\documentclass[12pt]{article}

\RequirePackage{amsthm,amsmath,amsfonts,amssymb}
\RequirePackage[colorlinks,citecolor=blue,urlcolor=blue]{hyperref}
\RequirePackage{graphicx}

\usepackage{float}
\usepackage{enumitem}
\usepackage{url}
\usepackage{bbm}
\usepackage{booktabs,caption}
\usepackage[flushleft]{threeparttable}
\usepackage{multirow}

\usepackage[pdftex,dvipsnames]{xcolor}
\usepackage{accents}
\usepackage{subfigure}
\usepackage{natbib}
\usepackage[margin=1in]{geometry}
\newcommand{\ba}{\boldsymbol{a}}
\newcommand{\bc}{\boldsymbol{c}}
\newcommand{\Cov}{\mathsf{Cov}}
\newcommand{\bD}{\boldsymbol{D}}
\newcommand{\E}{\mathsf{E}}
\newcommand{\bH}{\boldsymbol{H}}
\newcommand{\bI}{\boldsymbol{I}}
\newcommand{\bK}{\boldsymbol{K}}
\newcommand{\ols}{\mathsf{ols}}
\newcommand{\prob}{\mathsf{P}} 
\newcommand{\bU}{\boldsymbol{U}}
\newcommand{\V}{\mathsf{Var}}

\newcommand{\bX}{\boldsymbol{X}}
\newcommand{\by}{\boldsymbol{y}}
\newcommand{\bY}{\boldsymbol{Y}}
\newcommand{\bzero}{\boldsymbol{0}}
\newcommand{\bbeta}{\boldsymbol{\beta}}
\newcommand{\bdelta}{\boldsymbol{\delta}}
\newcommand{\bmu}{\boldsymbol{\mu}}
\newcommand{\bSigma}{\boldsymbol{\Sigma}}
\newcommand{\usefulness}{u}

\newcommand{\MSE}{\mathsf{MSE}}

\newtheorem{theorem}{Theorem}
\newtheorem{proposition}{Proposition}

\allowdisplaybreaks

\begin{document}

\title{\bf 
How useful is a wrong model? \\Information-sharing for inference under mean misspecification in linear models
}
\author{Emily C. Hector\hspace{.2cm}\\
Department of Biostatistics, University of Michigan}
\date{}
\maketitle
 
 \bigskip
\begin{abstract}
As almost all models are wrong, a mean model's usefulness is often accepted as sufficient justification for its use. In practice, however, standard statistical theory breaks when the mean model fit is imperfect, limiting this usefulness. This tension between fit and usefulness arises from a dichotomization of model fit: the model is either right or it is wrong. Motivated by the linear regression framework, we propose an alternative viewpoint  that leverages the mean model's usefulness \emph{without} assuming it is right or wrong. We define a new model usefulness index and use it to share information across individual observations through the mean model. The result is an estimator of each outcome's mean that shrinks individualized means towards the shared mean model, with the degree of shrinkage governed by this usefulness index. We draw connections between our estimator and the James-Stein estimator and establish when and how our estimators of the individualized means yield more efficient inference than model-based and non-model-based alternatives. We also propose a data-dependent estimate of the usefulness index that balances statistical intuition with efficiency considerations. We illustrate our method's practical value in an analysis of personal tracker data.
\end{abstract}

\noindent%
{\it Keywords:} Efficient inference, Goodness-of-fit, Information fusion, Penalized regression, Stein shrinkage.

\section{Introduction}
\label{s:intro}

The famous aphorism ``All models are wrong, but some are useful,'' usually attributed to George Box \citep{box1976science, box1987empirical, cox95discussion}, is often interpreted as license to use a model that makes sense for the data and gives useful insights, even though it does not fit the data perfectly. The ``usefulness'' of a model is essentially taken as sufficient justification for its use, despite it being ``wrong.'' Ignoring the imperfection of the model fit, however, can be dangerous: standard statistical theory breaks, and precious insights lose the theoretical support, such as type 1 error rate control, that makes them valuable. Standard post-fitting procedures to evaluate model fit are purely diagnostic, not prescriptive, and therefore do not on their own change the model to improve fit. 

A central limitation of this framing and one that gives rise to the tension between model fit and usefulness is that it dichotomizes model fit: the model is either wrong or it is right. To motivate our discussion, consider $n$ independent random variables $Y_i$, $i=1, \ldots, n$. We model $Y_i \sim \mathsf{N} (\mu_i, \sigma^2)$, with $\E(Y_i)=\mu_i^\star$ the true outcome mean. The central question considered in this paper is how to leverage the linear regression model $\mu_i=\bX_i^\top \bbeta$ when it is reasonable for some observations $y_i$ given covariates  $\bX_i \in \mathbb{R}^p$. The reality is that, often, the linear model is sufficient for the desired insights on some or even most of the observations in a sample. The model fits reasonably well to the bulk of the data in that it captures the general trend, but the model is imperfect or fits poorly to a few observations. One should generally question the suitability of trimming or Winsorizing outlying values, as is sometimes recommended \citep{wilcox2003applying}, or using robust regression methods, which down-weight these observations to estimate the regression parameter $\bbeta$ \citep{huber1964robust}, because the parameter $\bbeta$ loses its interpretability when used to describe data for which the model does not fit. Further, down-weighting or removing the contribution of some data points to the overall estimation inflates the variance of an estimator $\widehat{\bbeta}$, yet retaining even down-weighted data points for which the model does not fit introduces bias in the estimation of $\bbeta$. The statistician thus faces a catch-22: including such (down-weighted) data points introduces bias and challenges interpretability, while omitting them reduces efficiency. Efficiency loss (in a mean squared error sense) of the resulting model estimator $\bX_i^\top \widehat{\bbeta}$ of $\mu_i^\star$ therefore seems inevitable.

We show that loss of efficiency in estimation of $\mu_i^\star$ is not, in fact, inevitable in situations where a model is reasonable but not perfect. Our proposal, described in Section \ref{s:method}, is to move away from the now uninterpretable model parameter $\bbeta$ as a central estimand of interest and to focus instead on estimating $\mu_i^\star$. Our central philosophy does not dichotomize a model as either right or wrong but instead leverages how \emph{useful} it is in inferring $\mu_i^\star$. Our main idea, presented in Section \ref{s:estimation}, is to infer each outcome's mean by adaptively borrowing from the model according to its usefulness using a new \emph{usefulness index}. Some important connections to Stein shrinkage are also drawn in Section \ref{s:estimation}. In Section \ref{s:theory}, we show how to precisely quantify a model's usefulness by minimizing the mean squared error of the estimated $\mu_i^\star$. We show that the usefulness index depends on both how well the model fits (the model bias) and the variance of the outcome ($\sigma^2$). We also show that our estimators of $\mu_i^\star$ are more efficient than both usual estimators $y_i$ and $\bX_i^\top \widehat{\bbeta}_{\ols}$, where $\widehat{\bbeta}_{\ols}$ is the ordinary least squares estimator, and establish when and how to use our estimator for inference. As both the model bias and outcome variance are unknown and not jointly identifiable, Section \ref{s:implementation} discusses how to estimate the usefulness index. Section \ref{s:illustration} illustrates our method's practical value with an analysis of personal tracker data.

\section{A usefulness index for mean inference} \label{s:method}

\subsection{From model constraints to a usefulness index} \label{s:estimation}

We begin with a fresh perspective on the linear regression problem: we reformulate the likelihood by decoupling the mean model from other model components. This decoupling makes explicit the linear regression model's use of the assumption that the mean model is entirely useful. As our focus is on estimating $\bmu^\star=(\mu_1^\star, \ldots, \mu_n^\star) \in \mathbb{R}^n$, we view $\bmu=(\mu_1, \ldots, \mu_n)$ as the model parameters, and the mean structure $\mu_i=\bX_i^\top \bbeta$ as a \emph{constraint} on the model parameters. Given realizations $y_i$, maximum likelihood estimation can be rewritten as the constrained optimization problem of maximizing the log-likelihood (up to constants) $-\sum_{i=1}^n (y_i - \mu_i)^2$ subject to the constraint $\mu_i = \bX^\top_i \bbeta$, $i=1, \ldots, n$. The constrained solution is recovered as the limiting solution of the quadratic-penalty problem
\begin{align}
\arg \textstyle\max_{\bmu, \bbeta} \textstyle\sum_{i=1}^n \left\{ -(y_i - \mu_i)^2 - \usefulness ( \mu_i - \bX^\top_i \bbeta)^2 \right\},
\label{e:classic_obj}
\end{align}
where $\usefulness>0$ is a usefulness index: the hard-constraint formulation corresponds to the limit $\usefulness \to \infty$, in which the model is deemed entirely useful. In the limit $\usefulness \to \infty$, the solutions are the usual maximum likelihood estimators $\widehat{\bbeta}_{\ols}=(\bX^\top \bX)^{-1} \bX^\top \by$ and $\widehat{\mu}_i=\bX^\top_i \widehat{\bbeta}_{\ols}$, where $\by=(y_1, \ldots, y_n) \in \mathbb{R}^n$ and $\bX=(\bX_1, \ldots, \bX_n)^\top \in \mathbb{R}^{n \times p}$. 

When the model does not fit all observations, however, forcing the usefulness index to tend to $\infty$ is no longer justified as it may introduce bias in estimating $\mu_i^\star$. Nonetheless, the model may be useful, in the sense that we believe the means $\mu_i$ are relatively close to the structure $\bX_i^\top \bbeta$. This belief can be encoded by relaxing the hard constraint and allowing $\usefulness \geq 0$, not necessarily $\infty$, to be a usefulness index whose value is driven by how well the model $\bX_i^\top \bbeta$ fits $\mu_i$. Denote the maximizers in equation \eqref{e:classic_obj} by $\widehat{\bmu}(\usefulness), \widehat{\bbeta}(\usefulness)$. This objective function permits the borrowing of information in estimating $\bmu^\star$ based on how well the shared mean structure $\bX_i^\top \bbeta$ fits without committing fully to the mean model. We thus expect a gain in efficiency for our estimators of $\mu_i^\star$ relative to $y_i$, and a decrease in bias relative to the model-based estimator $\bX_i^\top \widehat{\bbeta}_{\ols}$. It remains to be seen whether both the variance and bias of $\widehat{\bmu}(\usefulness)$ can be sufficiently reduced to outperform both $\by$ and $\bX \widehat{\bbeta}_{\ols}$ in a mean squared error sense. 

The index $\usefulness$ must be specified by the user according to how useful they believe the model to be. Calling $\usefulness$ a usefulness index does not necessarily make it so; rather, how we choose $\usefulness$ will confer on this parameter the desired interpretation. Specifically, in Section \ref{s:theory} the usefulness index $\usefulness$ is chosen to minimize the mean squared error of $\widehat{\bmu}(\usefulness)$, explicitly tying usefulness to a reduction in mean squared error.

The estimator of $\mu_i^\star$ shrinks the outcomes $y_i$ through the relationship
\begin{align*}
\widehat{\bmu}(\usefulness)= \{ \bI_n + \usefulness ( \bI_n - \bH ) \}^{-1} \by,
\end{align*}
where $\bH = \bX (\bX^\top \bX)^{-1} \bX^\top \in \mathbb{R}^{n \times n}$ is the ``hat'' matrix (we show in Section \ref{s:theory} than $\bI_n - \bH$ is positive semi-definite). Using the Woodbury matrix identity and defining $\pi=(1+\usefulness)^{-1} \in [0,1]$, the estimator of $\bmu^\star$ can be rewritten as
\begin{align*}
\widehat{\bmu}(\pi) &= \pi \by + (1-\pi) \bH \by = \pi \by + (1-\pi) \bX \widehat{\bbeta}_{\ols} .
\end{align*}
Our estimator bears some similar to the best linear unbiased predictors from a random effects model, but without the restrictive distributional assumptions enforced by random effects modeling. Our estimator $\widehat{\bmu}(\pi)$ linearly interpolates between $\by$ and $\bX \widehat{\bbeta}_{\ols}$. When $\pi=1$ ($\usefulness=0$), the model is not useful and our estimator of $\bmu^\star$ reverts to the noisy but unbiased estimator $\by$; when $\pi=0$ ($\usefulness \to \infty$), the model is useful and our estimator of $\bmu^\star$ relies on the low variance but potentially biased linear model estimator $\bX \widehat{\bbeta}_{\ols}$. One can think of $\pi=1$ and $\pi=0$ as encoding situations where the model is respectively believed to be \emph{wrong} and \emph{right}. As discussed in Section \ref{s:intro}, however, our approach moves beyond this dichotomy by introducing a spectrum of model usefulness, encoded by $\pi \in (0,1)$, between these two extremes. In Section \ref{s:theory}, we formalize how to evaluate $\pi$, or equivalently the model usefulness index $\usefulness$, according to efficiency considerations. 

By moving away from the model parameter $\bbeta$ as the central estimand of interest and focusing instead on estimating $\bmu^\star$, our problem bears strong similarity to that considered by \cite{james1961estimation} and \cite{stein1981estimation}, and so it is not surprising that our philosophy has strong connections to the James-Stein estimator. As described in \cite{efron1973combining}, the James-Stein estimator combines estimation problems that are possibly related, where the possible relation is the proximity of the $\mu_i^\star$ to the origin. \cite{efron1973stein} couch the James-Stein estimator within a general class of empirical Bayes estimators that dominate the maximum likelihood estimator. Their proposal is to place a $\mathsf{N}(0, A)$ prior on $\mu_i$, where the centering at 0 naturally encodes the proximity to the origin of the James-Stein estimator as the shared relation between the $\mu_i$. Their proposed estimator, $\widehat{\bmu}=\{1 - \sigma^2/(\sigma^2+A)\} \by$, shrinks all outcomes towards the origin. The prior variance plays a role similar to the role of our usefulness index $\usefulness$ in that it tunes the shrinkage according to the prior belief in the similarity (i.e. proximity to the origin) of the estimation problems: $A\to \infty$ reverts to the noisy but unbiased estimator $\by$, and $A=0$ results in collapse of the estimator to the origin. In comparison, we propose the estimator $\widehat{\bmu}(\usefulness)=\{ (1+\usefulness)^{-1} \bI_n + \usefulness(1+\usefulness)^{-1} \bH \} \by$, which combines estimation problems that are possibly related by their proximity to the model $\bX_i^\top \bbeta$, rather than the origin. This is a more natural and useful relation when the model is believed to fit at least some observations.

Our perspective on the linear regression problem introduces a useful insight: the means are pooled together and estimated using a model with a shared regression parameter, $\bbeta$, that is common to all observations. The shared quantity $\bbeta$ permits the pooling, or integration, of information across observations. When estimating $\mu_i^\star$, this integration is fundamental to the efficiency of the estimator $\bX_i^\top \widehat{\bbeta}_{\ols}$ compared to the noisier estimator $y_i$. Our information-sharing approach can therefore be interpreted as a data integration method. Our integration philosophy bears some resemblance to that of \cite{hector2024turning} in transfer learning. Their problem is similar in the sense that both our context and the transfer learning setup aim to improve inference by borrowing from potentially useful but potentially harmful external information. In \cite{hector2024turning}, the parameter estimates of a target model are shrunk towards fixed parameter estimates from an independent source dataset, which constitute the external information; in our setup, the external information comes from the belief that the model is reasonable for some observations. Two important differences distinguish the two approaches, however: where the source estimates are fixed and source and target data sets are independent in the transfer learning problem, we estimate, in the same dataset, both the parameters that are being shrunk \emph{and} the parameter to which they shrink. 

Our approach provides substantial additional flexibility over the linear model in estimating $\mu_i^\star$, at no cost to interpretation when the model fits all observations. Indeed, our estimator of $\bbeta$ corresponds to the ordinary least squares estimator:
\begin{align*}
\widehat{\bbeta}(\usefulness)=(\bX^\top \bX)^{-1} \bX^\top \widehat{\bmu}(\usefulness) = \widehat{\bbeta}_{\ols}.
\end{align*}
The least squares estimator is intuitively appealing as it provides the line of best fit by minimizing the squared deviations from the model, $\sum_{i=1}^n (y_i - \bX_i^\top \bbeta)^2$, but \cite{buja2019models1, buja2019models2} note that the interpretation of $\bbeta$ changes under model misspecification. Our proposal offers both the usual line-of-best-fit interpretation an applied statistician may want and an inference for each observation's mean that acknowledges the model is wrong but potentially useful.

\subsection{Efficiency gains from evaluating model usefulness} \label{s:theory}

Our estimator $\widehat{\bmu}(\usefulness)$ lives on a spectrum between two estimators: the low bias, high variance estimator $\by$ and the high bias, low variance estimator $\bX \widehat{\bbeta}_{\ols}$. Our objective is to evaluate the usefulness of the model by finding a range of $\usefulness$ that minimizes both bias and variance and, ideally, outperforms both ends of the spectrum. The pseudo-true value of $\bbeta$, in the sense of \cite{white1982maximum}, is the root of the expected estimating function of $\bbeta$, $\bX^\top \bmu^\star - \bX^\top \bX \bbeta$, and is given by $\bbeta^\star = (\bX^\top \bX)^{-1} \bX^\top \bmu^\star$. The bias of the mean model, $\bdelta = (\delta_1, \ldots, \delta_n) = \bmu^\star - \bX \bbeta^\star$, is the difference between the outcome mean and the outcome mean model. Throughout, we use the notational convention $\ba^2 = \ba^\top \ba$ for a vector $\ba$. Occasionally, it will be simpler to work with $(\bdelta^2)^{-1}$; to allow situations in which $\bdelta^2 = 0$, we define its inverse through its limit: $(\bdelta^2)^{-1} = \infty$. Proofs of all results are given in Appendix \ref{a:proofs}. 

We first derive the mean squared error of $\widehat{\bmu}(\usefulness)$ in Theorem \ref{thm:gaussian-MSE}.
\begin{theorem}\label{thm:gaussian-MSE}
The mean squared error of $\widehat{\bmu}(\usefulness)$ is $\MSE\{ \widehat{\bmu}(\usefulness)\} = v(\usefulness) + b(\usefulness)$ with
\begin{align*}
v(\usefulness) &= (n-p) \sigma^2 (1+\usefulness)^{-2} +p \sigma^2, \quad b(\usefulness)= \usefulness^2(1+\usefulness)^{-2} \bdelta^2,
\end{align*}
respectively corresponding to the trace of the variance and the squared bias of $\widehat{\bmu}(\usefulness)$. 
\end{theorem}
When $\usefulness = 0$, $v(\usefulness) = n\sigma^2 = \mbox{trace} \{ \V (\by) \}$ and $b(\usefulness) = 0$, as expected. When $\usefulness \to +\infty$, $v(\usefulness) \to p\sigma^2 = \mbox{trace} \{ \V(\bX \widehat{\bbeta}_{\ols}) \}$ and $b(\usefulness) \to \bdelta^2$, also as expected. As the sample size $n$ grows, both the variance and bias diverge, suggesting that $\usefulness$ must be carefully chosen if inference on $\bmu^\star$ is to be informative. We will return to this point after Theorem \ref{thm:gaussian-opt} below.

Proposition \ref{thm:gaussian-naive} shows that there exists a wide range of $\usefulness$ over which our estimator outperforms the noisy but unbiased estimator $\by$.
\begin{proposition}\label{thm:gaussian-naive}
The mean squared error of $\widehat{\bmu}(\usefulness)$ is strictly smaller than the mean squared error of $\by$ for all $\usefulness>0$ when $\bdelta^2 \leq (n-p)\sigma^2$, and if and only if $0 <\usefulness < 2(n-p) \sigma^2/\{\bdelta^2 - (n-p)\sigma^2\}$ when $\bdelta^2 > (n-p)\sigma^2$.
\end{proposition}
When the outcome variance is small or the bias is large, Proposition \ref{thm:gaussian-naive} shows that estimation of $\bmu^\star$ can only borrow a small amount from the model $\bX \widehat{\bbeta}_{\ols}$ if it is to outperform $\by$; conversely, when the outcome variance is large or the bias is small, estimation of $\bmu^\star$ can borrow more from the model and still outperform $\by$. When $\sigma^2$ and $\bdelta^2$ are of the same order, more information can be safely borrowed from the model when the sample size is larger.

Next, Proposition \ref{thm:gaussian-model} establishes when our estimator outperforms the efficient by potentially biased estimator $\bX \widehat{\bbeta}_{\ols}$. 
\begin{proposition}\label{thm:gaussian-model}
The mean squared error of $\widehat{\bmu}(\usefulness)$ is strictly smaller than the mean squared error of $\bX \widehat{\bbeta}_{\ols}$ for all $\usefulness \geq 0$ when $\bdelta^2 > (n-p)\sigma^2$, and if and only if $\usefulness > (n-p) \sigma^2/(2\bdelta^2) - 1/2$ when $\bdelta^2 \leq (n-p)\sigma^2$.
\end{proposition}
The result of Proposition \ref{thm:gaussian-model} is complementary to that of Proposition \ref{thm:gaussian-naive}: when the outcome variance is large or the bias is small, estimation of $\bmu^\star$ must borrow more from the model if it is to outperform $\bX \widehat{\bbeta}_{\ols}$; conversely, when the outcome variance is small or the bias is large, estimation of $\bmu^\star$ can borrow less from the model $\bX \widehat{\bbeta}_{\ols}$ and still outperform $\bX \widehat{\bbeta}_{\ols}$. These results confirm our intuition that $\widehat{\bmu}(\usefulness)$ should align more closely to the model-based estimator when the bias is small or the outcome variance is large, and more closely to the outcome when the model bias is large or the outcome variance is small.

It remains to be seen if Proposition \ref{thm:gaussian-naive} and Proposition \ref{thm:gaussian-model} are compatible in the sense that $\usefulness$ can be chosen to outperform both ends of the spectrum simultaneously. Theorem \ref{thm:gaussian-opt} below establishes this, and something stronger as well: the usefulness index $\usefulness$ that minimizes the mean squared error of $\widehat{\bmu}(\usefulness)$ satisfies both inequalities of Propositions \ref{thm:gaussian-naive} and \ref{thm:gaussian-model}. 
\begin{theorem}\label{thm:gaussian-opt}
The mean squared error of $\widehat{\bmu}(\usefulness)$ is minimized at $\usefulness_\mathsf{or} = (n-p)\sigma^2 (\bdelta^2)^{-1}$ when $\bdelta^2 >0$, and at $\usefulness_\mathsf{or} \to \infty$ when $\bdelta^2=0$. Further, the mean squared error of $\widehat{\bmu}(\usefulness_\mathsf{or})$ is strictly smaller than the mean squared errors of both $\by$ and  $\bX \widehat{\bbeta}_{\ols}$ when $\bdelta^2>0$, and equal to that of $\bX \widehat{\bbeta}_{\ols}$ when $\bdelta^2=0$.
\end{theorem}
The usefulness of the model is determined by $\usefulness_\mathsf{or}=\sigma^2 \{ \bdelta^2/(n-p) \}^{-1}$, the size of the outcome variance relative to the average squared model bias. The mean model is more useful when the outcomes are noisy or when the model bias is small, and less useful when the model deviations are large relative to the noise. Importantly, model usefulness is not equivalent to correctness: from $\usefulness_\mathsf{or}>0$ when $\bdelta^2>0$, even a misspecified mean model is always useful. Although $\usefulness_\mathsf{or}$ depends on $n-p$, remember that $\bdelta^2=\sum_{i=1}^n \delta_i^2$ also typically grows linearly in $n$: model usefulness does not necessarily increase with sample size and instead depends on the ratio of the outcome variance to average squared model bias. 

Based on Theorem \ref{thm:gaussian-opt}, the oracle estimator of $\bmu$ is
\begin{align} \label{e:estimator}
\widehat{\bmu}_\mathsf{or}
&=\widehat{\bmu}(\usefulness_\mathsf{or}) =\frac{\bdelta^2}{\bdelta^2+(n-p)\sigma^2} \by + \frac{(n-p)\sigma^2}{\bdelta^2+(n-p)\sigma^2} \bX \widehat{\bbeta}_{\ols}. 
\end{align}
The form of $\widehat{\bmu}_\mathsf{or}$ does not require normality, but its mean squared error optimality does. The weights assigned to the naive and model-based estimators respectively correspond to the proportion of squared bias and (scaled) outcome variance, making explicit the bias-variance trade-off achieved by our formulation. The oracle estimator is unbiased at the two extremes: it corresponds to $\bX \widehat{\bbeta}_{\ols}$ when $\bdelta^2=0$, and it reverts to $\by$ when $\usefulness_\mathsf{or} \to 0$. 

The mean squared error of $\widehat{\bmu}_\mathsf{or}$ is
\begin{align*}
\frac{n\sigma^2 \bdelta^2 + p(n-p)\sigma^4}{\bdelta^2 + (n-p)\sigma^2} = p\sigma^2 + \frac{(n-p)\sigma^2 \bdelta^2}{\bdelta^2 +(n-p)\sigma^2}.
\end{align*}
When $\delta_i$ are allowed to depend on $n$, the mean squared error of $\widehat{\bmu}_\mathsf{or}$ converges to $p\sigma^2$ if $\limsup_{n \to \infty} \bdelta^2 = 0$, and it remains bounded below by $p\sigma^2$ if $0<\limsup_{n \to \infty} \bdelta^2 <\infty$. If $\limsup_{n \to \infty} \bdelta^2 = \infty$, the mean squared error of $\widehat{\bmu}_\mathsf{or}$ diverges; in particular, when $\bdelta^2$ is much larger than $n-p$, $\widehat{\bmu}_\mathsf{or}$ reverts to $\by$ and its mean squared error approaches $n \sigma^2$. To make this result formal, Proposition \ref{thm:gaussian-control} establishes necessary and sufficient conditions on $\bdelta^2$ for asymptotic (in $n$) control of the variance, bias and mean squared error of $\widehat{\bmu}_\mathsf{or}$. 
\begin{proposition}\label{thm:gaussian-control}
Suppose $\delta_i, i=1, \ldots, n,$ are allowed to depend on $n$.\\
As $n\to \infty$, the variance $v(\usefulness_\mathsf{or})$ of $\widehat{\bmu}_\mathsf{or}$ 
\begin{enumerate}\itemsep=0em
\item[](v.i) is bounded if and only if $\limsup_{n \to \infty} (n-p)^{-1/2}\sum_{i=1}^n \delta_i^2 < \infty$;
\item[](v.ii) converges to the variance of the minimum variance unbiased estimator under perfect model fit if and only if $\limsup_{n \to \infty} (n-p)^{-1/2} \sum_{i=1}^n \delta_i^2 = 0$. 
\end{enumerate}
As $n\to \infty$, the squared bias $b(\usefulness_\mathsf{or})$ of $\widehat{\bmu}_\mathsf{or}$ 
\begin{enumerate}
\item[](b.i) is bounded if and only if at least one of (a) $\limsup_{n \to \infty} (n-p)^{-2} \sum_{i=1}^n \delta_i^2 >0$ or (b) $\limsup_{n \to \infty} \sum_{i=1}^n \delta_i^2 < \infty$ holds;
\item[](b.ii) vanishes if and only if either (a) $\limsup_{n \to \infty} (n-p)^{-2} \sum_{i=1}^n \delta_i^2 = \infty$ or (b) $\limsup_{n \to \infty} \sum_{i=1}^n \delta_i^2 = 0$.
\end{enumerate}
As $n \to \infty$, the mean squared error of $\widehat{\bmu}_\mathsf{or}$
\begin{enumerate}
\item[](MSE.i) is bounded if and only if $\limsup_{n \to \infty} \sum_{i=1}^n \delta_i^2 < \infty$.
\item[](MSE.ii) converges to the variance of the minimum variance unbiased estimator under perfect model fit if and only if $\limsup_{n \to \infty} \sum_{i=1}^n \delta_i^2 = 0$.
\end{enumerate}
\end{proposition}

Conditions \textit{(b.i)} (respectively \textit{(b.ii)}) under which the bias of $\widehat{\bmu}_\mathsf{or}$ is controlled (resp. vanishes) as $n \to \infty$ may appear surprising: the squared model bias, $\sum_{i=1}^n \delta_i^2$, must either be asymptotically bounded (resp. vanish) or grow at a rate at least as fast (resp. faster) than $n^2$. The former makes sense -- if the squared model bias vanishes then $\bX \widehat{\bbeta}_{\ols}$ is asymptotically unbiased and any convex combination of $\bX \widehat{\bbeta}_{\ols}$ and $\by$ is also asymptotically unbiased -- but the latter may appear counter-intuitive. In fact, when $\lim_{n \to \infty} (n-p)^{-2} \sum_{i=1}^n \delta_i^2 = \infty$, 
\begin{align*}
\widehat{\bmu}_\mathsf{or} &=\frac{1}{1+(n-p)\sigma^2 (\bdelta^2)^{-1}} \by + \frac{1}{1 + (n-p)^{-1}\sigma^{-2}\bdelta^2} \bX \widehat{\bbeta}_{\ols} \to \by.
\end{align*}
In other words, $\usefulness_\mathsf{or}$ naturally adapts to situations in which the model bias grows by reverting to the noisy but unbiased estimator $\by$. This is important: the estimator protects against bias due to large misspecification by forcing no borrowing, but it does not protect against inefficiency. This suggests that the bias of $\widehat{\mu}_{\mathsf{or}, i}$
can be ignored when $\delta_i^2$ is small or when $(n-p)^{-1}\bdelta^2$ is large relative to $\sigma^2$. Under these conditions,
\begin{align} \label{e:large-ci}
&\widehat{\mu}_{\mathsf{or}, i} \pm z_{\alpha/2} \sigma \bigl\{ (1+\usefulness_\mathsf{or})^{-2} + h_{ii} -(1+\usefulness_\mathsf{or})^{-2} h_{ii} \bigr\}^{1/2}
\end{align}
is an approximate $100(1-\alpha)\%$ confidence interval for $\mu^\star_i$, where $h_{ii}$ is the $i$th diagonal element of $\bH$ and $z_{\alpha/2}$ is the $1-\alpha/2$ quantile of the standard normal distribution.

Of course, $\sigma^2$ and $\bdelta^2$ are unknown and must be estimated from the data in order to obtain $\widehat{\bmu}_\mathsf{or}$ in \eqref{e:estimator} and establish whether inference based on \eqref{e:large-ci} can be used. Section \ref{s:implementation} provides these details.

\subsection{Estimating the usefulness index in practice} \label{s:implementation}

In practice, we propose to find a data-driven version $\widehat{\usefulness}_\mathsf{or}$ of $\usefulness_\mathsf{or}$ from which we compute $\widehat{\bmu}=\widehat{\bmu}(\widehat{\usefulness}_\mathsf{or})$. Recall from Theorem \ref{thm:gaussian-opt} that $\usefulness_\mathsf{or} = (n-p) \sigma^2 (\bdelta^2)^{-1}$, and so estimates of both $\sigma^2$ and $\bdelta^2$ are needed.  When the mean model is correctly specified for all observations, $\sigma^2$ is estimated using the scaled sum of squared errors, $(n-p)^{-1} (\by - \bX \widehat{\bbeta}_{\ols} )^\top (\by - \bX \widehat{\bbeta}_{\ols} )$; when $\bdelta^2 >0$, however, this estimator is positively biased for $\sigma^2$:
\begin{equation} \label{e:delta_p_mean}
\begin{split}
&\E \bigl\{(n-p)^{-1} (\by - \bX \widehat{\bbeta}_{\ols} )^2 \bigr\}\\
&=(n-p)^{-1} \E \bigl\{ \by^\top (\bI_n-\bH) \by\bigr\}
=(n-p)^{-1}  \bigl[n \sigma^2 + \bmu^{\star\top} \bmu^\star - \mbox{trace}\{ \bH \E(\by \by^\top) \} \bigr]\\
&=(n-p)^{-1} [n\sigma^2 + \bmu^{\star\top} \bmu^\star - p\sigma^2 - \bmu^{\star\top} \bH \bmu^\star]
=\sigma^2 + (n-p)^{-1} \bmu^{\star\top} (\bI_n - \bH) \bmu^\star\\
&=\sigma^2 + (n-p)^{-1} \bdelta^2.
\end{split}
\end{equation}
On average, departures from the model are controlled by the squared model bias and the outcome variance. Indeed, both contribute to the individual squared deviations from the line of best fit through 
\begin{equation} \label{e:delta_i_mean}
\begin{split}
\E \{ (y_i - \bX_i^\top \widehat{\bbeta}_{\ols})^2 \} &= \sigma^2 + \sigma^2 \bX_i^\top (\bX^\top \bX)^{-1} \bX_i + \delta_i^2 - 2 \Cov(y_i, \bX_i^\top \widehat{\bbeta}_{\ols})\\
&= \sigma^2 (1- h_{ii} ) + \delta_i^2.
\end{split}
\end{equation}
Equations \eqref{e:delta_p_mean} and \eqref{e:delta_i_mean} show that the variance and squared bias terms, $\sigma^2$ and $\bdelta^2$, are not separately identifiable without additional information. But remember that our approach is motivated by the belief that the model fits at least some, if not most, observations reasonably well. This additional information can be used to develop a principled estimator of $\sigma^2$, which we denote by $\widehat{\sigma}_{(s)}^2$, described below. As the sum of squared errors is positively biased for $\bdelta^2$ with bias equal to $(n-p)\sigma^2$ (see equation \eqref{e:delta_p_mean}), we estimate $\bdelta^2$ with 
\begin{align*}
\widehat{\bdelta^2}= \max\{ (\by - \bX \widehat{\bbeta}_{\ols} )^2 - (n-p) \widehat{\sigma}^2_{(s)}, 0\}.
\end{align*}
From there, plugging estimators of $\sigma^2$ and $\bdelta^2$ into the form of $\usefulness_\mathsf{or}$ given in Theorem \ref{thm:gaussian-opt}, the empirical version of $\usefulness_\mathsf{or}$ is $\widehat{\usefulness}_\mathsf{or} = (n-p) \widehat{\sigma}_{(s)}^2 (\widehat{\bdelta^2})^{-1}$.

We now describe our estimator $\widehat{\sigma}_{(s)}^2$ of $\sigma^2$. Rearranging equation \eqref{e:delta_i_mean}, the estimators $\widehat{\sigma}_i^2 = (y_i - \bX_i^\top \widehat{\bbeta}_{\ols})^2 (1- h_{ii})^{-1}$ are positively biased estimators of $\sigma^2$, with bias $\delta_i^2 (1- h_{ii})^{-1}$, $i=1, \ldots, n$. As the investigator believes that the model $\bX_i^\top \bbeta$ is reasonable for at least some outcome means, the bias of $\widehat{\sigma}_i^2$ should be small for at least some $i \in \{1, \ldots, n\}$. We propose to estimate $\sigma^2$ using the $s$th quantile of the $\widehat{\sigma}_i^2$, denoted by $\widehat{\sigma}_{(s)}^2$, where $s$ must be chosen by the investigator. For example, an investigator concerned that the model may only fit very few outcomes may take a more conservative approach by selecting a smaller order statistic. This conservative estimator of $\sigma^2$ would lead to a conservative estimator of $\usefulness_\mathsf{or}$ that favours $\by$ over the mean model. In general, we recommend investigating diagnostic plots to select $s$, and $s$ can usually be chosen by eyeballing quantile-quantile plots of sample residuals from the linear model fit -- see the simulation below and the data illustration in Section \ref{s:illustration}. 

It is worth emphasizing that our estimator of $\sigma^2$ is new in the context of Stein shrinkage. For example, \cite{james1961estimation} assume $\sigma^2$ is known or that a statistic with mean $\sigma^2$ independent of $\by$ is available. In their sequence of papers, \cite{efron1973combining, efron1973stein, efron1975data} advocate for a Bayesian approach that places a prior on $\mu_i$, where choosing the prior variance reflects the level of heterogeneity in the $\mu_i$. This approach restricts the level of heterogeneity in the model fit that can be modeled. Indeed, while both our and these papers' approaches require the user to pre-specify how well they think the model fits, we do not require the deviations from the model to follow any particular model or distribution. 

We can derive a more general form of the James-Stein estimator, one that shrinks towards $\bX \widehat{\bbeta}_{\ols}$ instead of the origin, following the arguments in Section 5 (see equation (4.21)) of \cite{lehmann1998theory}: 
\begin{align}
\label{e:James-Stein}
\widehat{\bmu}_{\mathsf{JS}}&= \bX \widehat{\bbeta}_{\ols} + \Bigl\{ 1 - \frac{(n-p-2)\sigma^2}{(\by - \bX \widehat{\bbeta}_{\ols} )^2} \Bigr\} ( \by - \bX \widehat{\bbeta}_{\ols}).
\end{align}
Interestingly, this estimator corresponds to $\widehat{\bmu}(\widehat{\usefulness}_\mathsf{JS})$ with $\widehat{\usefulness}_\mathsf{JS}=(n-p-2)\sigma^2\{(\by - \bX \widehat{\bbeta}_{\ols})^2 - (n-p-2)\sigma^2\}^{-1}$. The modified James-Stein estimator is therefore a version of our approach with a slight modification of the usefulness index. The only real difference between $\widehat{\bmu}_{\mathsf{JS}}$ and $\widehat{\bmu}$, apart from assuming the variance is known, is the SURE correction factor $-2$. When the variance is known and we assume that $(\by - \bX \widehat{\bbeta}_{\ols})^2 > (n-p)\sigma^2$, we can also derive a SURE-type correction for $\widehat{\bmu}$ that accounts for the additional variability due to estimating $\usefulness_\mathsf{or}$; it replaces the factor $n-p$ by $n-p-2$, reducing our estimator exactly to the modified James-Stein estimator $\widehat{\bmu}_\mathsf{JS}$! 

\begin{figure}[ht]
\centering
\includegraphics[width=\textwidth]{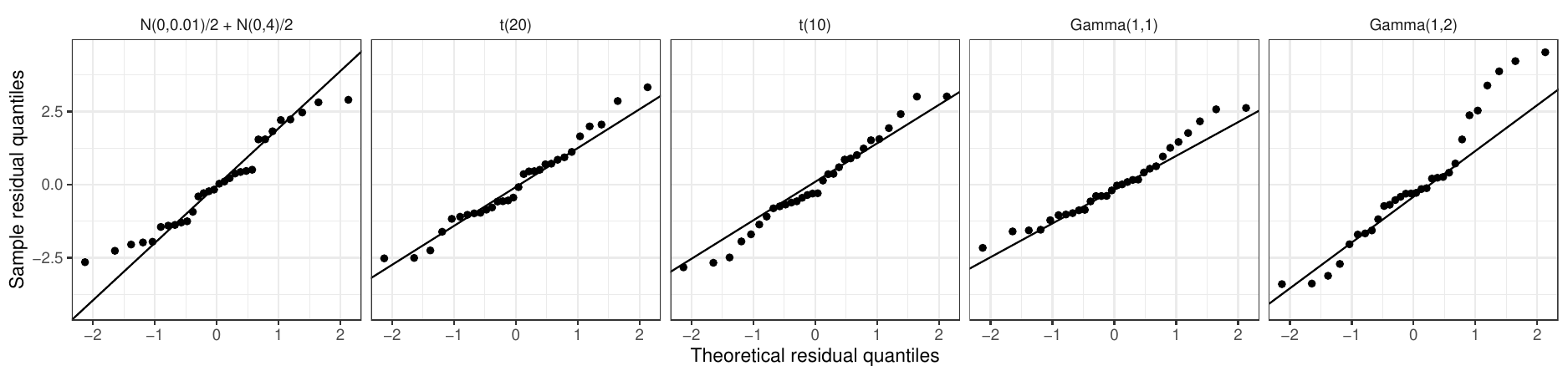}
\caption{Quantile-quantile plot of the residuals from the linear model fit on one representative dataset in the five simulation settings. \label{f:simulations-qq}}
\end{figure}

Ideally, we would show that $\widehat{\bmu}$ enjoys properties similar to those established for $\widehat{\bmu}_\mathsf{or}$ in Section \ref{s:theory}, but this is especially difficult because of the form of $\widehat{\usefulness}_\mathsf{or}$'s dependence on the data. Instead, we empirically evaluate the performance of $\widehat{\bmu}$ in a simulation. We consider a setup with $n=30$ observations and $p=6$ covariates consisting of an intercept and five continuous variables. We set $\bmu^\star = \bX \bbeta + \bc$, with continuous covariates independently generated from standard normal distributions, $\bbeta=(1.33, 0.9, 1.9, -0.13, 1.25, -1.18)$ and $\bc=(c_1, \ldots, c_n)$ specified next; as $\bdelta = (\bI_n-\bH) \bmu^\star$, this definition of $\bmu^\star$ gives $\bdelta = (\bI_n-\bH) \bc$. The $c_i$ are independently generated from one of five distributions: (i) an equally weighted mixture of mean-zero normal distributions with variances $0.01$ and $4$, $\mathsf{N}(0,0.01)/2 + \mathsf{N}(0,4)/2$; (ii) a $t$ distribution with 20 degrees of freedom, $\mathsf{t}(20)$; (iii) a $t$ distribution with 10 degrees of freedom, $\mathsf{t}(10)$; (iv) a Gamma distribution with shape and scale set to 1, $\mathsf{Gamma}(1,1)$; and (v) a Gamma distribution with shape and scale set to 1 and 2 respectively, $\mathsf{Gamma}(1,2)$. In each of 1000 Monte Carlo replicates, outcomes $y_i$ are independently generated from a normal distribution with mean $\bmu^\star$ and variance $\sigma^2=1$. Figure \ref{f:simulations-qq} 
displays a quantile-quantile plot of the residuals from the linear model fit on one representative dataset in each of the five settings. These plots show evidence of mild to severe lack of fit for several observations, motivating the use of an inference based on model usefulness. Across settings (i)-(iv), the model seems reasonable for at least half of the observations, and so we set $s=n/2$. In setting (v), model fit seems adequate for only approximately a third of the data points; to illustrate the behaviour of $\widehat{\bmu}$ when $s$ is too large, we initially set $s=n/2$ in setting (v) as well, with $s=n/3$ investigated subsequently.

We compare the oracle estimator $\widehat{\bmu}_\mathsf{or}$ and the proposed estimator $\widehat{\bmu}$, with $\widehat{\usefulness}_\mathsf{or}$ computed using $\widehat{\sigma}^2_{(n/2)}$, to the estimators $\bX \widehat{\bbeta}_{\ols}$ and $\by$, and to the modified James-Stein estimator defined in equation \eqref{e:James-Stein} with known variance. Empirical mean squared errors (and their standard errors) computed from the 1000 Monte Carlo replicates are reported in Table \ref{t:simulation-MSE}. 
\begin{table}[ht]
\centering
\begin{tabular}{rlrrrrr}
\multicolumn{2}{l}{Setting} & $\bX \widehat{\bbeta}_{\ols}$ & $\by$ & $\widehat{\bmu}_{\mathsf{JS}}$ & $\widehat{\bmu}_\mathsf{or}$ & $\widehat{\bmu}$\\
  \hline\\[-0.9em]
(i) & $\frac{\mathsf{N}(0,0.01)}{2} + \frac{\mathsf{N}(0,4)}{2}$& 37.2 (1.06) & 30.0 (2.40) & 20.6 (1.65) & 19.5 (1.54) & 20.6 (1.62) \\ 
(ii) & $\mathsf{t}(20)$ & 26.1 (1.06) & 30.0 (2.40) & 18.3 (1.48) & 16.9 (1.33) & 18.0 (1.49) \\ 
(iii) & $\mathsf{t}(10)$ & 32.9 (1.06) & 30.0 (2.40) & 19.9 (1.55) & 18.7 (1.44) & 20.2 (1.53) \\ 
(iv) & $\mathsf{Gamma}(1,1)$ & 41.5 (1.06) & 30.0 (2.40) & 21.3 (1.69) & 20.2 (1.60) & 21.7 (1.73) \\ 
(v) & $\mathsf{Gamma}(1,2)$ & 148 (1.06) & 30.0 (2.40) & 27.0 (2.17) & 26.5 (2.13) & 43.4 (4.78) \\ 
\end{tabular}
\caption{Empirical mean squared errors (and their standard errors$\times 10$) computed from 1000 Monte Carlo replicates. \label{t:simulation-MSE}}
\end{table}
As guaranteed by the theory, the mean squared error of $\widehat{\bmu}_\mathsf{or}$ is always smaller than that of $\bX \widehat{\bbeta}_{\ols}$ and $\by$. Its mean squared error is also always smaller than that of $\widehat{\bmu}_\mathsf{JS}$. In settings (i)-(iv), the mean squared error of $\widehat{\bmu}$ tracks that of $\widehat{\bmu}_\mathsf{or}$ closely, suggesting that our data-driven $\widehat{\usefulness}_\mathsf{or}$ is reasonably close to $\usefulness_\mathsf{or}$. The estimator $\widehat{\bmu}$ also performs nearly identically to the modified James-Stein estimator with known variance in settings (i)-(iv). In setting (v), using $\widehat{\sigma}^2_{(n/2)}$ overestimates $\sigma^2$ and underestimates $\bdelta^2$, and $\widehat{\usefulness}_\mathsf{or}$ therefore overestimates $\usefulness_\mathsf{or}$. Figure \ref{f:simulations-qq} shows, however, that the model fit is poor for more than half of observations; indeed, model fit seems adequate for only approximately a third of the data points. The proposed estimator $\widehat{\bmu}$ with $\widehat{\usefulness}_\mathsf{or}$ computed using $\widehat{\sigma}^2_{(n/3)}$ instead has mean squared error $27.4$ (standard error $0.227$), which not only outperforms the mean squared error of $\bX \widehat{\bbeta}_{\ols}$ and $\by$ but is also much closer to the mean squared error of the oracle estimator $\widehat{\bmu}_\mathsf{or}$; it also matches closely to the mean squared error of the modified James-Stein estimator with known variance. 

\begin{figure}
\centering
\includegraphics[width=\textwidth]{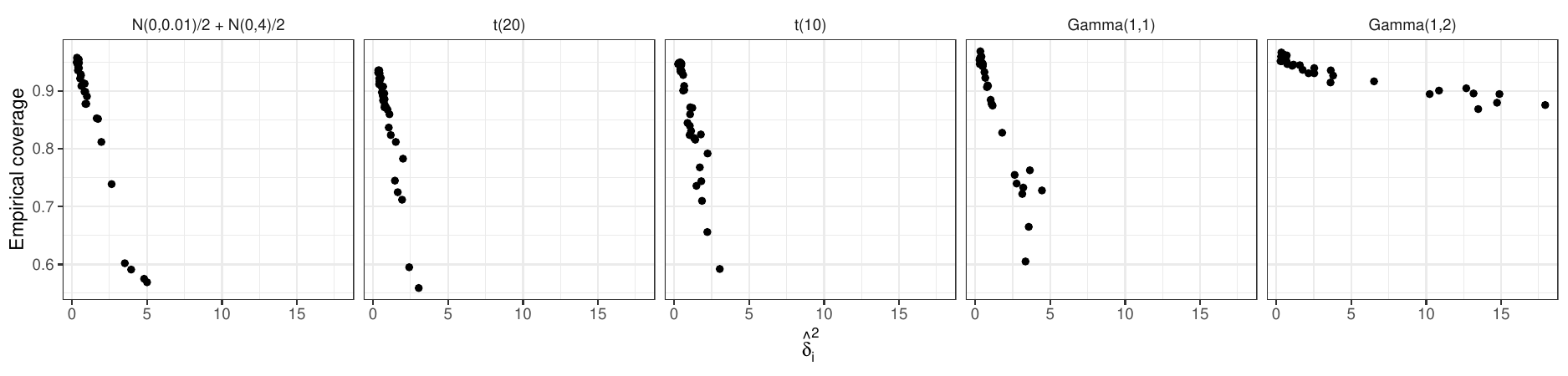}
\caption{Monte Carlo average of $\widehat{\delta}_i^2$ versus empirical coverage of 95\% confidence intervals. \label{f:coverage-plot}}
\end{figure}

Figure \ref{f:coverage-plot} compares, for each observation, the Monte Carlo average of the estimated squared model bias,
\begin{align*}
\widehat{\delta}_i^2=\max\{(y_i - \bX_i^\top \widehat{\bbeta}_{\ols})^2 - \widehat{\sigma}^2_{(n/2)} (1-h_{ii}), 0\}
\end{align*}
derived from equation \eqref{e:delta_i_mean}, with the empirical coverage of the corresponding 95\% confidence interval $\widehat{\mu}_i \pm z_{\alpha/2} \widehat{\sigma}_{(s)} \{ (1+\widehat{\usefulness}_\mathsf{or})^{-2} + h_{ii} - (1+\widehat{\usefulness}_\mathsf{or})^{-2} h_{ii} \}^{1/2}$. We use $s=n/2$ in settings (i)-(iv) and $s=n/3$ in setting (v). 
As expected, the coverage deteriorates as $\widehat{\delta}_i^2$ increases in settings (i)-(iv): when an individual mean is poorly approximated by the fitted model and the fitted model is given sufficient weight by $\widehat{\usefulness}_\mathsf{or}$, the bias introduced by the shrinkage is no longer negligible. In setting (v), however, coverage remains close to the nominal level for large values of $\widehat{\delta}_i^2$. This occurs because the overall squared model bias is much larger in setting (v), causing $\widehat{\usefulness}_\mathsf{or}$ to favor the estimator $\by$ and thereby reducing shrinkage-induced bias. Specifically, $(n-p)^{-1}\bdelta^2$ takes values $1.31$, $0.846$, $1.13$, $1.49$ and $5.94$ in settings (i)-(v) respectively: as $(n-p)^{-1}\bdelta^2$ is much larger than $\sigma^2=1$ in setting (v), the estimated $\widehat{\usefulness}_\mathsf{or}$ downweights the model. Moreover, the corresponding Monte Carlo averages of $(n-p)^{-1} \widehat{\bdelta^2}$ are $1.59$, $1.22$, $1.36$, $1.69$ and $5.99$, suggesting that $(n-p)^{-1} \widehat{\bdelta^2}$ is a reasonable estimator of squared model bias. Together, these results provide some guidance on when to expect confidence intervals to approximately reach nominal levels: large estimated local bias $\widehat{\delta}_i^2$ can harm coverage when the estimator still borrows substantially from the model, but has less impact when $(n-p)^{-1} \widehat{\bdelta^2}$ is large relative to $\widehat{\sigma}^2_{(s)}$ because the estimated usefulness index appropriately downweights the model. 

\section{Illustration} \label{s:illustration}

We illustrate our method's practical value with an analysis of data from $n=33$ Fitbit users who responded to a survey distributed via Amazon Mechanical Turk \citep{furberg_2016_53894}. Users submitted personal tracker data from April 12 to May 12, 2016. Personal trackers such as Fitbits and Apple watches provide a unique opportunity for personalized insights into how physical activity improves human health. On the one hand, well established relationships between physical activity and calories burned have been established for general populations that can help guide and inform clinical interventions. On the other hand, some individuals find their metabolism responds uniquely to physical activity and the general published guidelines fail to capture their unique caloric response to physical activity. Using general guidelines to set physical activity goals for these individuals may yield sub-optimal clinical outcomes. What is needed for these individuals is an inference that, while informed by the association between physical activity and caloric response of other study participants, is uniquely shaped to their metabolism.

\begin{figure}[ht]
\centering
\subfigure[Observed vs fitted outcomes.]{\includegraphics[width=0.47\textwidth]{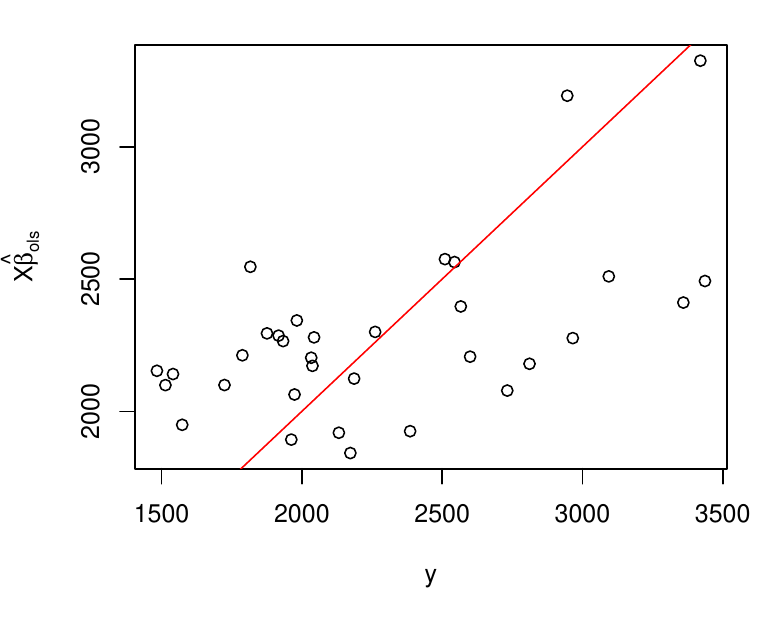}}
\subfigure[Fitted outcomes vs standardized residuals.]{\includegraphics[width=0.47\textwidth]{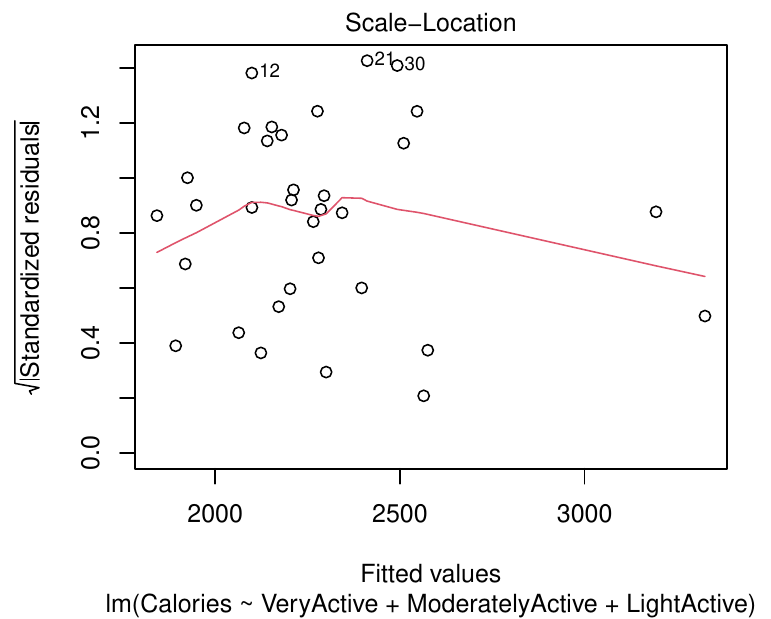}}
\subfigure[Theoretical quantiles vs standardized residuals.]{\includegraphics[width=0.47\textwidth]{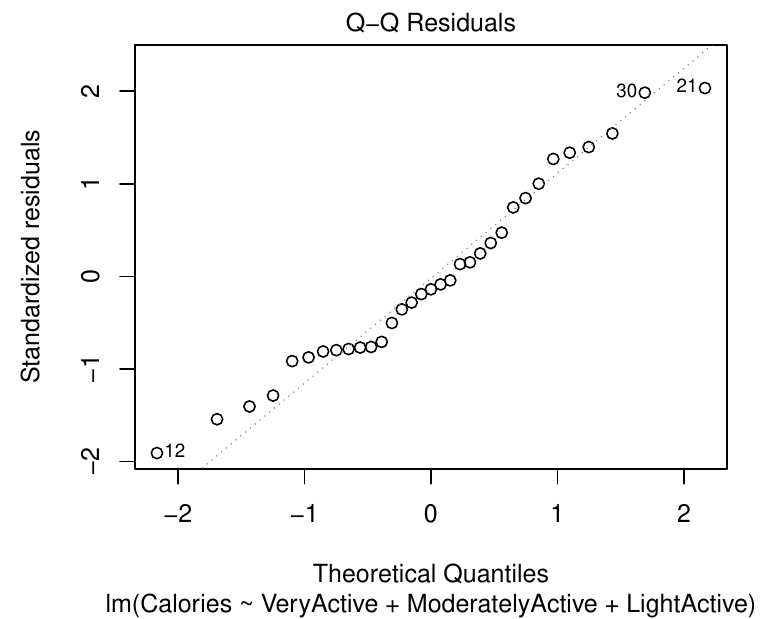}}
\subfigure[Leverage vs standardized residuals.]{\includegraphics[width=0.47\textwidth]{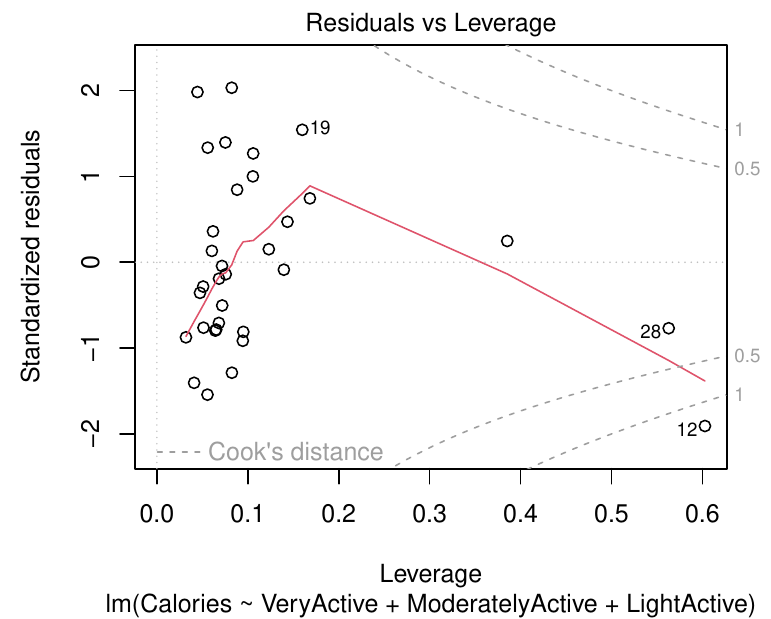}}
\caption{Diagnostic plots for linear model fit. \label{f:diagnostic-plots}}
\end{figure}

We model the average daily calories burned (over the period from April 12 to May 12) $y_i$ as a function of covariates $\bX_i$ consisting of an intercept ($x_{i1}$) and the average of daily distance in ``very active'' ($x_{i2}$), ``moderately active'' ($x_{i3}$) and ``light active'' ($x_{i4}$) efforts.  Marginal correlations ($p$-value) between the outcome and $\bX_2, \bX_3, \bX_4$ are respectively $0.5116$ ($0.0023$), $0.06198$ ($0.73$) and $0.3511$ ($0.045$), suggesting that these covariates explain some but not all of the variation in the outcome. The results of the linear model fit are reported in Table \ref{t:linear-model-fit}. The proportion of variance explained by the model is $R^2=0.3214$, and its adjusted counterpart is $R^2_{\text{adj}}=0.2513$, suggesting serviceable fit for drawing scientific insights. The overall regression $F$-test is significant, with $F=4.579$ on 3 and 29 degrees of freedom, and a $p$-value of $0.00961$, indicating that the activity variables jointly explain a statistically significant amount of variation in average daily calories burned. The moderate value of $R^2_\text{adj}$, however, together with the standard diagnostic plots in Figure \ref{f:diagnostic-plots}, suggest that the fitted mean structure captures only part of the heterogeneity across participants.
\begin{table}[ht]
\centering
\begin{tabular}{rrrrr}
Coefficients & Estimate & Std. Error & $t$-value & $\prob(>\mid t \mid)$    \\
\midrule
Intercept & 1778.01 & 227.41 & 7.818 & $1.27 \times 10^{-8}$ *** \\
Very Active & 139.01 & 47.89 & 2.903 & 0.00699 **  \\
Moderately Active & $-$85.25 & 166.88 & $-$0.511 & 0.61330   \\ 
Light Active & 105.63 & 66.41 & 1.591 & 0.12254  
\end{tabular}
\caption{Output from linear model fit: coefficient estimates, standard errors, $t$ statistic value and $p$-value. Stars indicate the strength of significance (*** for $p$-value$<0.001$, ** for $p$-value$<0.01$). \label{t:linear-model-fit}}
\end{table}

Based on the diagnostic plots in Figure \ref{f:diagnostic-plots}, the model appears to fit reasonably well to most observations, with a few noticeable exceptions. Indeed, observations $i \in \{12, 19, 21, 30\}$ are high leverage observations with large residuals. The usual estimator of $\sigma^2$ based on the standardized sum of squared errors is $\widehat{\sigma}^2=(n-p)^{-1} (\by - \bX \widehat{\bbeta}_{\ols})^2=487^2$, whereas our estimator is $\widehat{\sigma}^2_{(n/2)}=382^2$. This difference is important. The usual residual mean square combines both outcome variability and lack of fit of the mean structure, whereas $\widehat{\sigma}^2_{(n/2)}$ is intended to isolate the former. The gap between these estimates reflects variation that is better interpreted as mean misspecification rather than irreducible outcome variance. 

Figure \ref{f:continuum-fit} in Appendix \ref{a:supp-illustration} plots $\widehat{\bmu}(\usefulness)$ and $\widehat{\bmu}(\pi)$ for sequences of $\usefulness$ and $\pi$. The data-driven $\widehat{\usefulness}_\mathsf{or}$ using $\widehat{\sigma}^2_{(n/2)}$ and its counterpart $\widehat{\pi}_\mathsf{or}=(1+\widehat{\usefulness}_\mathsf{or})^{-1}$ are also shown there using dashed vertical lines. The estimated usefulness index $\widehat{\usefulness}_\mathsf{or} = 1.60$ indicates that the fitted mean model is useful but far from perfect. Equivalently, $\widehat{\pi}_\mathsf{or} = 0.385$, so the proposed estimator assigns weight $0.385$ to the individual outcomes $\by$ and weight $0.615$ to the fitted mean model $\bX \widehat{\bbeta}_{\ols}$. Thus, the data-driven procedure treats the linear model as carrying substantial information about individual caloric response, but not enough information to justify full reliance on the model-based estimator. 

\begin{figure}[h!]
\includegraphics[width=\textwidth]{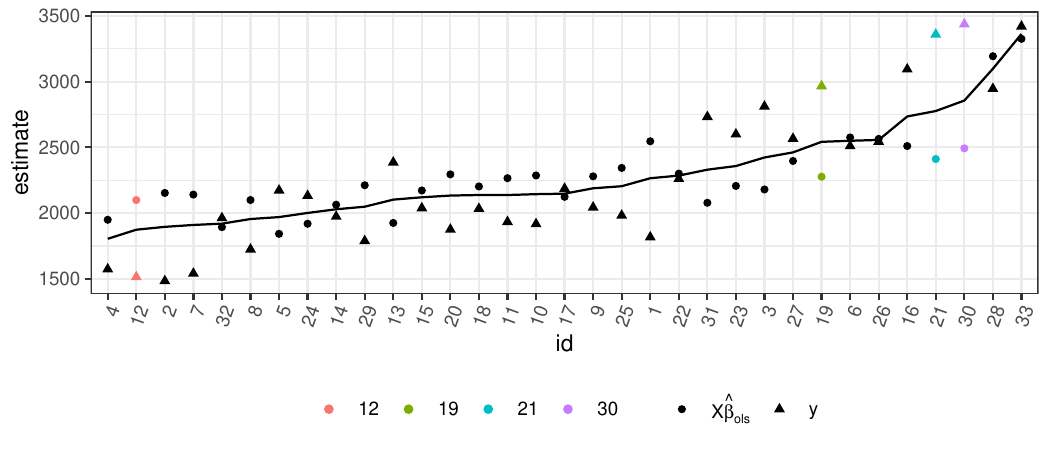}
\caption{Noisy ($y_i$) and biased ($\bX_i^\top \widehat{\bbeta}_{\ols}$) point estimates, with line denoting estimates ($\widehat{\bmu}$) for all participants, with specific individuals highlighted. Participants are sorted from smallest to largest outcome values. \label{f:comparison-grouped}}
\end{figure}

The linear model pulls the more extreme observations towards a central estimate that pools information from all observations equally. The linear model estimate $\bX_i^\top \widehat{\bbeta}_{\ols}$ is thus too large for small observations (e.g. $i=12$) and too small for large observations (e.g. $i \in \{ 19,21,30\}$). Estimates $\widehat{\bmu}$ are plotted in Figure \ref{f:comparison-grouped} along with estimates $\bX \widehat{\bbeta}_{\ols}$ and $\by$.
Observations 12, 19, 21 and 30 have noticeably large residuals. These are precisely the participants for whom the model may be inappropriate, and these observations illustrate the practical value of the usefulness index. For these individuals, the estimated caloric response differs meaningfully from what would be estimated based only on the shared activity-calorie relationship. Our method does not classify these observations as either ``well fit'' or ``outliers'' to be discarded; instead, it uses the estimated usefulness of the model to determine how strongly each participant should be pulled toward the shared regression fit. Further, the proposed method allows these individuals to influence inference without ignoring the information contained in the broader sample. This is especially relevant in personal tracker applications, where the scientific goal is often not only to estimate an average relationship but also to identify when individual responses deviate from that relationship.

\begin{figure}[ht]
\centering
\includegraphics[width=\textwidth]{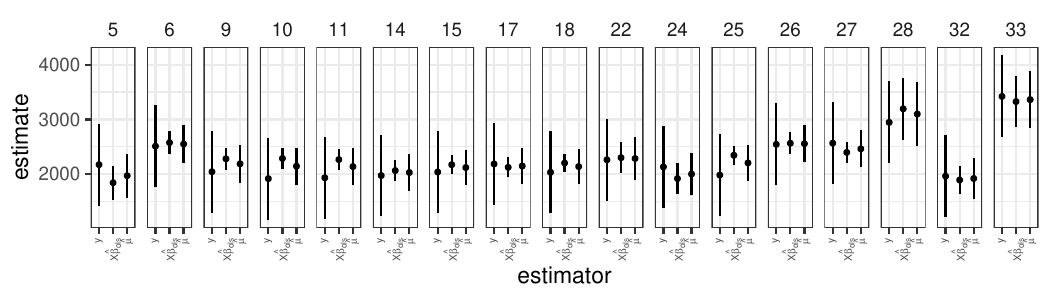}
\caption{95\% confidence intervals for $\bmu^\star$ using estimators $\by$, $\bX \widehat{\bbeta}_{\ols}$ and $\widehat{\bmu}$.} \label{f:comparison-CI}
\end{figure}

The estimate of $(n-p)^{-1} \bdelta^2$ is $(n-p)^{-1} \widehat{\bdelta^2}=91334.9$, which is smaller than $\widehat{\sigma}^2_{(n/2)}=382^2$. We therefore build 95\% confidence intervals using $\widehat{\mu}_i \pm z_{\alpha/2} \widehat{\sigma}_{(n/2)} \{ (1+\widehat{\usefulness}_\mathsf{or})^{-2} + h_{ii} - (1+\widehat{\usefulness}_\mathsf{or})^{-2} h_{ii} \}^{1/2}$ only for participants with $\widehat{\delta}_i^2=0$. These are plotted in Figure \ref{f:comparison-CI} along 95\% confidence intervals based on $\bX_i^\top \widehat{\bbeta}_{\ols} \pm z_{\alpha/2} \widehat{\sigma}_{(n/2)} h_{ii}^{1/2}$ and $y_i \pm z_{\alpha/2} \widehat{\sigma}_{(n/2)}$.
Intervals based on $\by$ are wide because they borrow no information across participants, and intervals based on $\bX_i^\top \widehat{\bbeta}_{\ols}$ are narrow because they fully pool information through the regression model at the cost of potential bias. The proposed estimator lies between these extremes: the resulting intervals are therefore narrower than those based on $\by$, and their centers less exposed to bias than those based on $\bX_i^\top \widehat{\bbeta}_{\ols}$. The linear mean model is not fully correct for all participants, but it remains useful for improving inference about individual outcome means.

\section{Discussion}

The goal of this paper is not to determine whether a mean model is right or wrong, but rather to determine how useful a wrong mean model is for inferring outcome means from a normal model. Our usefulness index $\usefulness_\mathsf{or}$ and its data-driven counterpart $\widehat{\usefulness}_\mathsf{or}$ provide clear guidance: a wrong mean model is useful, in the sense that it reduces mean squared error, when the outcomes are noisy relative to the average squared model bias. The usefulness index measures how much information should be shared between observations through the fitted mean model. A model is also always useful, in a mean squared error sense, even when it is incorrect, because the usefulness index automatically balances the variance reduction with the bias introduced by the mean model. The proposed estimator adapts naturally to lack of model fit by up or downweighting the fitted mean model according to its estimated bias. Inference for the individual mean estimates is reliable when the estimated local bias $\widehat{\delta}_i^2$ is small, or when the estimated standardized squared model bias is large relative to the estimated outcome variance, causing the estimator to borrow less from the model. 

An essential practical challenge is that $\sigma^2$ and $\bdelta^2$ are not separately identifiable without additional information. Our key idea is to leverage external information, namely that the model is reasonable for some subset of the observations, to derive a quantile-based estimator of $\sigma^2$. This is intentionally guided by the user, rather than fully automatic: the choice of quantile reflects the user's skepticism about mean model fit. A smaller quantile yields a more conservative estimate of model usefulness and therefore less borrowing from the fitted model. This conservative approach results in a larger variance of $\widehat{\bmu}$ but protects against bias. 

The proposed estimator $\widehat{\bmu}$ is closely related to a modified James-Stein estimator that combines possibly related estimation problems, where the possible relation is proximity to the fitted model $\bX \widehat{\bbeta}_{\ols}$ rather than the origin. Under simplifying assumptions, the connection to James-Stein reveals the familiar SURE correction. An exact correction for the data-driven estimator $\widehat{\bmu}$, however, is not tractable because $\widehat{\usefulness}_\mathsf{or}$ depends on the data through a quantile-based variance estimate and a truncated estimate of $\bdelta^2$. The simulations suggest that, despite this complication, the data-driven usefulness index performs well when the quantile is chosen sensibly. 

The global usefulness index $\usefulness$ provides a clean interpretation on how information is shared across individuals, but it also imposes a common degree of borrowing across observations. Individual-specific usefulness indices $\usefulness_i$ would permit greater flexibility for outlying and central mean values to shrink by different amounts. This flexibility is attractive, especially in applications where model fit varies substantially across individuals, but it introduces a much harder identification problem because one needs to separate $\sigma^2$ and individual-specific squared biases $\delta_i^2$. Given the challenging nature of separately identifying $\sigma^2$ and $\bdelta^2$ with the global usefulness index, reliable estimation of individual-specific usefulness indices seems unlikely.

This paper focuses on inference for $\mu^\star_i$ given the availability of both $y_i$ and $\bX_i^\top \widehat{\bbeta}_{\ols}$. A related goal is out-of-sample prediction of a new observation $y_{n+1}$, assumed to follow $\mathsf{N} (\mu^\star_{n+1}, \sigma^2)$, given covariates $\bX_{n+1}$. While our estimator reduces to the model-based predictor $\bX_{n+1}^\top \widehat{\bbeta}_{\ols}$ in the prediction setting due to the unavailability of $y_{n+1}$, the usefulness index nonetheless remains informative: it quantifies the reliability of the model-based predictor by measuring the average gap between the fitted mean structure and the true outcome means.

\section*{Acknowledgments}

This work was supported by a grant from the National Science Foundation (DMS~2337943). The author is grateful for insightful conversations with Dr.  Ryan Martin.

\appendix

\section{Proofs} \label{a:proofs}

\subsection{Proof of Theorem \ref{thm:gaussian-MSE}}

Denote $\V(\by)=\sigma^2 \bI_n$ and let $\bK=\bI_n-\bH$. By some simple algebra, $\bK \bmu^\star = \bmu^\star - \bH \bmu^\star = \bmu^\star - \bX \bbeta^\star$. Using this,
\begin{align*}
&\{ \widehat{\bmu}(\usefulness) - \bmu^\star \}^\top \{ \widehat{\bmu}(\usefulness) - \bmu^\star \} \\
&= \{ \by - ( \bI_n + \usefulness\bK) \bmu^\star \}^\top ( \bI_n + \usefulness \bK)^{-2} \{ \by - ( \bI_n + \usefulness\bK) \bmu^\star \}\\
&=\{ \by - \bmu^\star - \usefulness (\bmu^\star - \bX \bbeta^\star) \}^\top ( \bI_n + \usefulness \bK)^{-2} \{ \by - \bmu^\star - \usefulness (\bmu^\star - \bX \bbeta^\star) \}.
\end{align*}
Let $\bSigma = \sigma^2 \bI_n$. The mean squared error of the estimator is therefore
\begin{align*}
\MSE\{ \widehat{\bmu}(\usefulness)\} &= \E [\{ \widehat{\bmu}(\usefulness) - \bmu^\star \}^\top \{ \widehat{\bmu}(\usefulness) - \bmu^\star \} ]\\
&=\mbox{trace} \{ \bSigma (\bI_n + \usefulness \bK)^{-2} \} + \usefulness^2(\bmu^\star - \bX \bbeta^\star)^{\top} (\bI_n +\usefulness \bK)^{-2} (\bmu^\star - \bX \bbeta^\star)\\
&=\mbox{trace} \{ \bSigma (\bI_n + \usefulness \bK)^{-2} \} + \usefulness^2 \mbox{trace} \{(\bI_n + \usefulness \bK)^{-2} \bdelta \bdelta^{\top} \}\\
&=v(\usefulness) + b(\usefulness).
\end{align*}
The term $v(\usefulness)$ is the variance of the estimator and the term $b(\usefulness)$ is the bias term and will be $0$ when $\usefulness=0$ or $\bmu^\star = \bX \bbeta^\star$. Clearly, $v(\usefulness)$ is monotonically decreasing and $b(\usefulness)$ is monotonically increasing. Note that, as expected,
\begin{align*}
v(0) + b(0) = \mbox{trace} ( \bSigma \bI_n^{-2} ) = \MSE(\by),
\end{align*}
the mean squared error of the na\"{i}ve estimator of the mean $\bmu^\star$. When $\bmu^\star = \bX \bbeta^\star$, we therefore have $\MSE\{\widehat{\bmu}(\usefulness)\} < \MSE (\by)$ for all $\usefulness>0$. Throughout the remainder of the proof, we consider the case when $\bmu^\star \neq \bX \bbeta^\star$. 

We first simplify the variance term $v(\usefulness)$. The matrix $\bH$ is idempotent, as $\bH \bH = \bH$, and therefore all its eigenvalues are either zero or one. As the largest eigenvalue of $\bH$ is less than or equal to one, the matrix $\bK = \bI_n - \bH$ is positive semi-definite. Notice that $\bH$ is positive semi-definite with $\mbox{trace}(\bH) = \mbox{trace}\{ (\bX^\top \bX)^{-1} \bX^\top \bX \}=\mbox{trace}(\bI_p)=p$. As the trace is equal to the sum of the eigenvalues, $\bH$ has $p$ eigenvalues equal to one and $n-p$ eigenvalues equal to $0$.

The matrix $\bK$ is symmetric and positive semi-definite and so it admits a decomposition of the form $\bK=\bU^\top \bD \bU$ with $\bU^{-1}=\bU^\top$, $\bU$ the matrix of eigenvectors of $\bK$ and $\bD=\mbox{diag}\{ \eta_i(\bK)\}_{i=1}^n$ the diagonal matrix of eigenvalues of $\bK$ sorted in increasing order, $0 \leq \eta_1(\bK) \leq \ldots \leq \eta_n(\bK)$. 
Therefore,
\begin{align*}
\bI_n + \usefulness \bK &= \bI_n + \usefulness \bU^\top \bD \bU = \bU^\top (\bI_n + \usefulness \bD ) \bU.
\end{align*}
For $\usefulness>0$, the matrix $\bI_n + \usefulness \bD$ is symmetric, positive definite and invertible. That is,
\begin{align*}
(\bI_n + \usefulness \bK)^{-2} &= [ \{\bU^\top ( \bI_n + \usefulness \bD ) \bU\}^{-1} ]^\top \{\bU^\top ( \bI_n + \usefulness \bD ) \bU\}^{-1}\\
&= \bU^\top ( \bI_n + \usefulness \bD)^{-2} \bU.
\end{align*}
Further, as $\bK=\bI_n-\bH$ is idempotent with trace $n-p$, it has eigenvalues $\eta_1(\bK)=\ldots=\eta_p(\bK)=0$ and $\eta_{p+1}(\bK)=\ldots=\eta_n(\bK)=1$. The variance term of the mean squared error simplifies to
\begin{align*}
v(\usefulness) = \sigma^2 \mbox{trace} \{ ( \bI_n + \usefulness \bD)^{-2} \} = \sigma^2 \textstyle \sum_{i=1}^n \{ 1 + \usefulness \eta_i(\bK) \}^{-2}= (n-p) \sigma^2 (1+\usefulness)^{-2} + p\sigma^2.
\end{align*}

We now simplify the bias term, $b(\usefulness)$, in the mean squared error of $\widehat{\bmu}(\usefulness)$. Using the Woodbury matrix identity,
\begin{align*}
(\bI_n + \usefulness \bK)^{-1}&= (1+\usefulness)^{-1} \bI_n + \usefulness(1+\usefulness)^{-1} \bH = (1+\usefulness)^{-1}(\bI_n + \usefulness \bH).
\end{align*}
Substituting this into $b(\usefulness)$,
\begin{align*}
b(\usefulness) 
&= \usefulness^2 (1+\usefulness)^{-2} \mbox{trace} \{ (\bI_n + \usefulness \bH)^2 \bdelta \bdelta^\top \}.
\end{align*}
As $\bH$ is idempotent, $(\bI_n + \usefulness \bH)^2 = \bI_n + \usefulness(2+\usefulness)\bH$. Plugging this into the form of the bias term,
\begin{align*}
b(\usefulness) &= \usefulness^2 (1+\usefulness)^{-2} \mbox{trace} \{ ( \bI_n + \usefulness(2+\usefulness)\bH) \bdelta \bdelta^\top \}\\
&=\usefulness^2(1+\usefulness)^{-2} \bdelta^2 + \usefulness^3(1+\usefulness)^{-2} (2+\usefulness) \mbox{trace} (\bH \bdelta \bdelta^\top ).
\end{align*}
Now, notice that $\bH \bdelta = \bH \bmu^\star - \bH \bX \bbeta^\star = \bX \bbeta^\star - \bX \bbeta^\star = \bzero$. Therefore, the bias term further simplifies to
\begin{align*}
b(\usefulness) &= \usefulness^2(1+\usefulness)^{-2} \bdelta^2.
\end{align*}
Putting the variance and bias pieces together, the mean squared error of $\widehat{\bmu}(\usefulness)$ is
\begin{align*}
\MSE\{ \widehat{\bmu}(\usefulness)\} &= v(\usefulness) + b(\usefulness) = (n-p) \sigma^2 (1+\usefulness)^{-2} +p \sigma^2 + \usefulness^2(1+\usefulness)^{-2} \bdelta^2.
\end{align*}

\subsection{Proof of Proposition \ref{thm:gaussian-naive}}

We show that there exists a range of $\usefulness>0$ such that $\MSE\{ \widehat{\bmu}(\usefulness)\} < \MSE(\bY)$, i.e.
\begin{align*}
(n-p) \sigma^2 (1+\usefulness)^{-2} +p \sigma^2 + \usefulness^2(1+\usefulness)^{-2} \bdelta^2 < n\sigma^2.
\end{align*}
Rearranging the above inequality, $\MSE\{ \widehat{\bmu}(\usefulness)\} < \MSE(\bY)$ if and only if
\begin{align*}
\usefulness \Bigl\{ \frac{\bdelta^2}{(n-p) \sigma^2} -1 \Bigr\} < 2.
\end{align*}
When $\bdelta^2 \leq (n-p)\sigma^2$, the bracketed term is non-positive and the above inequality holds for all $\usefulness>0$. When $\bdelta^2 > (n-p)\sigma^2$, the bracketed term is strictly positive, in which case $\MSE\{ \widehat{\bmu}(\usefulness)\} < \MSE(\bY)$ if and only if
\begin{align*}
0 <\usefulness < \frac{2(n-p) \sigma^2} {\bdelta^2 - (n-p)\sigma^2}.
\end{align*}

\subsection{Proof of Proposition \ref{thm:gaussian-model}}

Using the fact that $\bX \widehat{\bbeta}_{\ols} = \bH \by$,
\begin{align*}
&( \bX \widehat{\bbeta}_{\ols} - \bmu^\star )^\top ( \bX \widehat{\bbeta}_{\ols} - \bmu^\star )\\
&= ( \bH \by -\bH \bmu^\star + \bH \bmu^\star - \bmu^\star)^\top (\bH \by -\bH \bmu^\star + \bH \bmu^\star - \bmu^\star)\\
&=\{ \bH (\by- \bmu^\star) - \bK \bmu^\star \}^\top \{ \bH (\by- \bmu^\star) - \bK \bmu^\star \}\\
&=(\by - \bmu^\star)^\top \bH (\by-\bmu^\star) - (\by - \bmu^\star)^\top \bH \bK \bmu^\star - \bmu^{\star\top} \bK \bH (\by - \bmu^\star) + \bmu^{\star\top} \bK \bmu^\star\\
&=(\by - \bmu^\star)^\top \bH (\by-\bmu^\star) + \bmu^{\star\top} \bK \bmu^\star.
\end{align*}
Further, $\bK \bmu^\star = \bmu^\star - \bH \bmu^\star = \bmu^\star - \bX \bbeta^\star$. Let $\bSigma = \sigma^2 \bI_n$. The mean squared error of $\bX \widehat{\bbeta}_{\ols}$ is therefore
\begin{align*}
\MSE ( \bX \widehat{\bbeta}_{\ols}) &= \E \{ (\bX \widehat{\bbeta}_{\ols} - \bmu^\star )^\top ( \bX \widehat{\bbeta}_{\ols} - \bmu^\star ) \}\\
&=\mbox{trace} (\bH \bSigma) + \bmu^{\star\top} \bK \bmu^\star\\
&=p\sigma^2 + (\bK \bmu^\star)^2\\
&=p\sigma^2 + \bdelta^2,
\end{align*}
where we have used the fact that $\bK$ is symmetric and idempotent in the penultimate line of the above display. 

The mean squared error of $\widehat{\bmu}(\usefulness)$ is
\begin{align*}
\MSE\{ \widehat{\bmu}(\usefulness)\} 
&= (n-p) \sigma^2 (1+\usefulness)^{-2} +p \sigma^2 + \usefulness^2(1+\usefulness)^{-2} \bdelta^2.
\end{align*}
Therefore, $\widehat{\bmu}(\usefulness)$ has a smaller mean squared error than $\bX \widehat{\bbeta}_{\ols}$ if and only if
\begin{align*}
(n-p) \sigma^2 + \usefulness^2 \bdelta^2 < (1+\usefulness)^2 \bdelta^2.
\end{align*}
Equivalently, after some algebra, $\widehat{\bmu}(\usefulness)$ has a smaller mean squared error than $\bX \widehat{\bbeta}_{\ols}$ if and only if
\begin{align*}
\usefulness > \frac{(n-p) \sigma^2}{2\bdelta^2} - \frac{1}{2}.
\end{align*}
In the special case that $\bdelta^2 > (n-p)\sigma^2$, the lower bound in the right-hand side of the above display is negative, and $\widehat{\bmu}(\usefulness)$ has a smaller mean squared error than $\bX \widehat{\bbeta}_{\ols}$ for all values of $\usefulness \geq 0$.

\subsection{Proof of Theorem \ref{thm:gaussian-opt}}

The minimizer of $\MSE\{\widehat{\bmu}(\usefulness)\}$ corresponds to the root of its derivative,
\begin{align*}
\frac{\partial}{\partial \usefulness} \MSE\{\widehat{\bmu}(\usefulness)\} &= \frac{\partial}{\partial \usefulness} v(\usefulness) + \frac{\partial}{\partial \usefulness} b(\usefulness) = -2(n-p)\sigma^2(1+\usefulness)^{-3} + 2\usefulness(1+\usefulness)^{-3} \bdelta^2.
\end{align*}
If $\bdelta^2=0$, the root is obtained in the limit as $\usefulness \to \infty$, and $\widehat{\bmu}(\usefulness_\mathsf{or})=\bX \widehat{\bbeta}_{\ols}$. In the remainder of the proof, we consider the case where $\bdelta^2>0$. The root of $\partial \MSE\{\widehat{\bmu}(\usefulness)\} / (\partial \usefulness)$ is
\begin{align*}
\usefulness_\mathsf{or} &= (n-p)\sigma^2 (\bdelta^2)^{-1}.
\end{align*}
We consider each of the two possible configurations of the relative size of bias and variance separately, and show that in each configuration the mean squared error of $\widehat{\bmu}(\usefulness_\mathsf{or})$ is strictly smaller than the mean squared error of both $\by$ and $\bX \widehat{\bbeta}_{\ols}$. 

We first consider the configuration when $\bdelta^2 > (n-p)\sigma^2$. In this case, the mean squared error of $\widehat{\bmu}(\usefulness)$ is strictly smaller than the mean squared error of $\bX \widehat{\bbeta}_{\ols}$ for all $\usefulness \geq 0$, and strictly smaller than the mean squared error of $\by$ if and only if $0 <\usefulness < 2(n-p) \sigma^2/\{\bdelta^2 - (n-p)\sigma^2\}$. By definition,
\begin{align*}
\usefulness_\mathsf{or} = \frac{(n-p)\sigma^2}{\bdelta^2} < \frac{(n-p)\sigma^2}{ \bdelta^2 - (n-p)\sigma^2} < \frac{2(n-p)\sigma^2}{ \bdelta^2 - (n-p)\sigma^2}.
\end{align*}
Therefore, $\MSE\{ \widehat{\bmu}(\usefulness_\mathsf{or})\} < \MSE(\bY)$ and $\MSE\{ \widehat{\bmu}(\usefulness_\mathsf{or})\} < \MSE ( \bX \widehat{\bbeta}_{\ols} )$ when $\bdelta^2 > (n-p)\sigma^2$.

We next consider the configuration when $(n-p)\sigma^2 \geq \bdelta^2$. In this case, the mean squared error of $\widehat{\bmu}(\usefulness)$ is strictly smaller than the mean squared error of $\by$ for all $\usefulness>0$, and strictly smaller than the mean squared error of $\bX \widehat{\bbeta}_{\ols}$ if and only if $\usefulness > (n-p) \sigma^2/(2\bdelta^2) - 1/2$. By definition,
\begin{align*}
\usefulness_\mathsf{or} = \frac{(n-p)\sigma^2}{\bdelta^2} > \frac{(n-p)\sigma^2}{2\bdelta^2} > \frac{(n-p)\sigma^2}{2\bdelta^2} -\frac{1}{2}.
\end{align*}
Therefore, $\MSE\{ \widehat{\bmu}(\usefulness_\mathsf{or})\} < \MSE(\bY)$ and $\MSE\{ \widehat{\bmu}(\usefulness_\mathsf{or})\} < \MSE ( \bX \widehat{\bbeta}_{\ols} )$ when $(n-p)\sigma^2 \geq \bdelta^2$.

\subsection{Proof of Proposition \ref{thm:gaussian-control}}

Using Theorem \ref{thm:gaussian-MSE}, the variance of $\widehat{\bmu}_\mathsf{or} $ is
\begin{align*}
v(\usefulness_\mathsf{or}) &= p \sigma^2 + \frac{(n-p) \sigma^2}{\{ 1 + (n-p)\sigma^2 (\bdelta^2)^{-1} \}^2}\\
&=p \sigma^2 + \frac{1}{(n-p)^{-1} \sigma^{-2} + (n-p) \sigma^2 (\bdelta^2)^{-2} + 2 (\bdelta^2)^{-1}}.
\end{align*}
The variance $v(\usefulness_\mathsf{or})$ is controlled as $n \to \infty$ if and only if $\limsup_{n \to \infty} (n-p)^{-1/2}\sum_{i=1}^n \delta_i^2 < \infty$. Further, $v(\usefulness_\mathsf{or}) \to p\sigma^2$, the variance of the minimum variance unbiased estimator under perfect model fit ($\bdelta = \bzero$), if and only if $\limsup_{n \to \infty} (n-p)^{-1/2} \sum_{i=1}^n \delta_i^2 = 0$.

The squared bias is 
\begin{align*}
b(\usefulness_\mathsf{or}) 
&= \frac{(n-p)^2\sigma^4 (\bdelta^2)^{-1}}{\{ 1+(n-p)\sigma^2 (\bdelta^2)^{-1}\}^2}\\
&=\frac{1}{(n-p)^{-2} \sigma^{-4} (\bdelta^2) + (\bdelta^2)^{-1} + 2 (n-p)^{-1} \sigma^{-2}}.
\end{align*}
The squared bias $b(\usefulness_\mathsf{or})$ is controlled as $n \to \infty$ when either $\limsup_{n \to \infty} (n-p)^{-2} \sum_{i=1}^n \delta_i^2 >0$ or $\limsup_{n \to \infty} \sum_{i=1}^n \delta_i^2 < \infty$. Further, $b(\usefulness_\mathsf{or})\to 0$ when either $\limsup_{n \to \infty} (n-p)^{-2} \sum_{i=1}^n \delta_i^2 = \infty$ or $\limsup_{n \to \infty} \sum_{i=1}^n \delta_i^2 = 0$.

The variance and the squared bias are asymptotically controlled if and only if both $\limsup_{n \to \infty} (n-p)^{-1/2}\sum_{i=1}^n \delta_i^2 < \infty$ and one of $\limsup_{n \to \infty} (n-p)^{-2} \sum_{i=1}^n \delta_i^2 >0$ or \\ $\limsup_{n \to \infty} \sum_{i=1}^n \delta_i^2 < \infty$ are true. As $\limsup_{n \to \infty} \sum_{i=1}^n \delta_i^2 < \infty$ implies $\limsup_{n \to \infty} (n-p)^{-1/2}\sum_{i=1}^n \delta_i^2 < \infty$, the variance and squared bias are asymptotically controlled if and only if $\limsup_{n \to \infty} \sum_{i=1}^n \delta_i^2 < \infty$ or $\limsup_{n \to \infty} (n-p)^{-1/2} \sum_{i=1}^n \delta_i^2 < \infty$ and $\limsup_{n \to \infty} (n-p)^{-2} \sum_{i=1}^n \delta_i^2 >0$. Notice that one cannot have both $\limsup_{n \to \infty} (n-p)^{-1/2} \sum_{i=1}^n \delta_i^2 < \infty$ and $\limsup_{n \to \infty} (n-p)^{-2} \sum_{i=1}^n \delta_i^2 >0$. Therefore, the variance and squared bias are asymptotically controlled if and only if $\limsup_{n \to \infty} \sum_{i=1}^n \delta_i^2 < \infty$. 

Further, the mean squared error converges to $p\sigma^2$ if and only if both $\limsup_{n \to \infty} (n-p)^{-1/2} \sum_{i=1}^n \delta_i^2 = 0$ and one of $\limsup_{n \to \infty} (n-p)^{-2} \sum_{i=1}^n \delta_i^2 = \infty$ or $\limsup_{n \to \infty} \sum_{i=1}^n \delta_i^2 = 0$ are true. Notice that one cannot have both $\limsup_{n \to \infty} (n-p)^{-1/2} \sum_{i=1}^n \delta_i^2 = 0$ and $\limsup_{n \to \infty} (n-p)^{-2} \sum_{i=1}^n \delta_i^2 = \infty$. Therefore, the mean squared error converges to $p\sigma^2$ if and only if $\limsup_{n \to \infty} \sum_{i=1}^n \delta_i^2 = 0$.

\section{Additional illustration details} \label{a:supp-illustration}

\begin{figure}[ht]
\centering
\subfigure[Estimated means as a function of $\usefulness$.]{\includegraphics[width=0.45\textwidth]{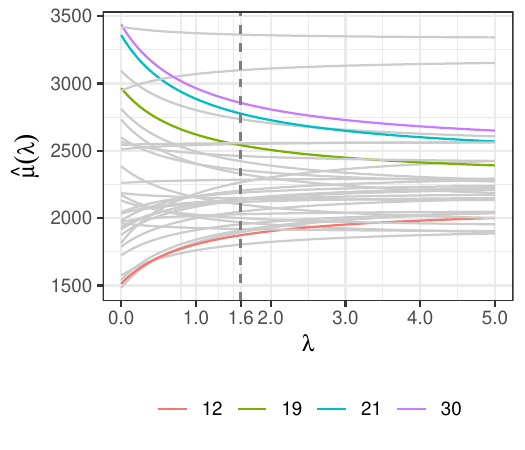}}
\subfigure[Estimated means as a function of $\pi$.]{\includegraphics[width=0.45\textwidth]{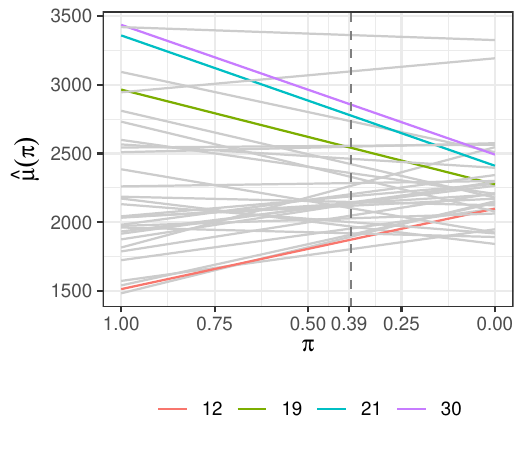}}
\caption{Estimated means as a function of shrinkage parameters for all participants, with specific individuals highlighted. Horizontal dashed lines denote values of the data-driven $\widehat{\usefulness}_\mathsf{or}$ and $\widehat{\pi}_\star$ in plots (a) and (b) respectively. \label{f:continuum-fit}}
\end{figure}

Figure \ref{f:continuum-fit} plots $\widehat{\bmu}(\usefulness)$ and $\widehat{\bmu}(\pi)$ for sequences of $\usefulness$ and $\pi$ in the illustration of Section \ref{s:illustration}. 

\bibliography{IF-bib-20240506}

\end{document}